\documentclass[twocolumn,showpacs,preprintnumbers,amsmath,amssymb]{revtex4-2}
\usepackage{graphicx}
\usepackage{dcolumn}
\usepackage{bm}
\usepackage{lineno}
\usepackage{comment}
\newcommand{\bef}{\begin{figure}}
\newcommand{\eef}{\end{figure}}

\newcommand{\be}{\begin{equation}}
\newcommand{\ee}{\end{equation}}
\newcommand{\bea}{\begin{eqnarray}}
\newcommand{\eea}{\end{eqnarray}}
\begin{document}

\title{Dipolar flow of identified hadrons at mid-rapidity using transport models}

\author{Abhisek Praharaj, Aradhana Panday, Sandeep Chatterjee, Md Nasim}
\affiliation{Department of Physical Sciences, Indian Institute of Science Education and Research, Berhampur, India}

\begin{abstract}
We report a transport model study of the rapidity even component of dipolar flow, $v_{1}^{\mathrm{even}}$, for identified charged hadrons at mid-rapidity in Au+Au collisions at $\sqrt{s_{NN}} = 7.7-200$ GeV. The analysis is performed using the AMPT model. In addition, HIJING model is used to quantify residual non-flow contributions in $v_{1}^{\mathrm{even}}$.
The $v_{1}^{\mathrm{even}}$ of identified hadrons ($\pi$, $K$, and $p$) shows no significant difference between particles and anti-particles at $\sqrt{s_{NN}} = 200$ GeV. However, a clear splitting between proton and anti-proton $v_{1}^{\mathrm{even}}$ develops with decreasing beam energy, while no corresponding difference is observed for mesons ($\pi^{\pm}$ and $K^{\pm}$). 
These results indicate that the proton-antiproton difference in $v_{1}^{\mathrm{even}}$ is sensitive to baryon transport and antibaryon annihilation in the hadronic medium. Our study highlights the potential of identified-particle $v_{1}^{\mathrm{even}}$ measurements at RHIC Beam Energy Scan energies as a novel probe of baryon stopping and the evolution of the hadronic medium.
\end{abstract}
\maketitle

\section{Introduction}

The study of collective behavior in relativistic heavy-ion collisions provides crucial insights into the properties of the strongly interacting matter created under extreme conditions~\cite{flow1,flow2,flow3,flow4}. Among various observables, anisotropic flow coefficients have proven to be powerful tools for probing the initial geometry, transport properties, and dynamical evolution of the system. 
``Flow" observables, characterized by the coefficients  in the Fourier expansion of azimuthal 
distributions of produced particle is given by
\begin{equation}
    v_{n} =\langle \rm{cos}\ n(\phi-\psi) \rangle.
\end{equation}

In this definition, $\phi$ and $\psi$ denote the azimuthal angles of an outgoing particle and the event reaction plane, 
respectively~\cite{flow_method}. The directed flow, characterized by the first-order Fourier coefficient $v_1$, is sensitive to the early-time dynamics of the collision and the subsequent expansion of the medium~\cite{v1_odd_0,v1_odd_1,v1_odd_2,v1_odd_3,v1_odd_4,v1_odd_5}.

Traditionally, directed flow is dominated by a rapidity-odd component arising from the collective sideward deflection of particles. However, fluctuations in the initial state can cause dipole-like asymmetries in the initial geometry giving rise to rapidity-even component of directed flow ($v_1^{\mathrm{even}}$)~\cite{v3_paper,v1_even_0,v1_even_1,v1_even_2,v1_even_3,v1_even_4}. This rapidity-even directed flow has attracted significant attention in recent years, as it provides complementary information about initial-state fluctuations and medium response. Previous studies have reported measurements of rapidity-even directed flow for charged hadrons, establishing its dependence on transverse momentum and collision centrality~\cite{v1_even_2,v1_even_3,v1_even_4}. Recent studies based on hydrodynamical models show $v_1^{\mathrm{even}}$ of identified hadrons can be a sensitive probe for baryon stopping and diffusion over rapidity~\cite{v1e_p_ap_hyd}. In this work, we extend the studies of $v_1^{\mathrm{even}}$ of identified hadrons using transport based models. 
The $v_1^{\mathrm{even}}$ of identified hadrons can be sensitive the transported-quark mechanism , annihilation in hadronic medium, hadronic and partonic mean-field potentials in baryon-rich matter~\cite{trans_1,trans_2,anhilation_1,mean_field_1,mean_field_2,mean_field_3,mean_field_4}. 
 In this study, we employ two widely used transport models: HIJING (Heavy Ion Jet INteraction Generator)~\cite{hijing} and AMPT (A Multi-Phase Transport Model)~\cite{{ampt}}. While HIJING primarily incorporates initial-state particle production and jet physics, AMPT includes partonic and hadronic interactions, making it well-suited for studying the development of collective behavior.

\begin{figure*}[ht]
\includegraphics[scale=0.8]{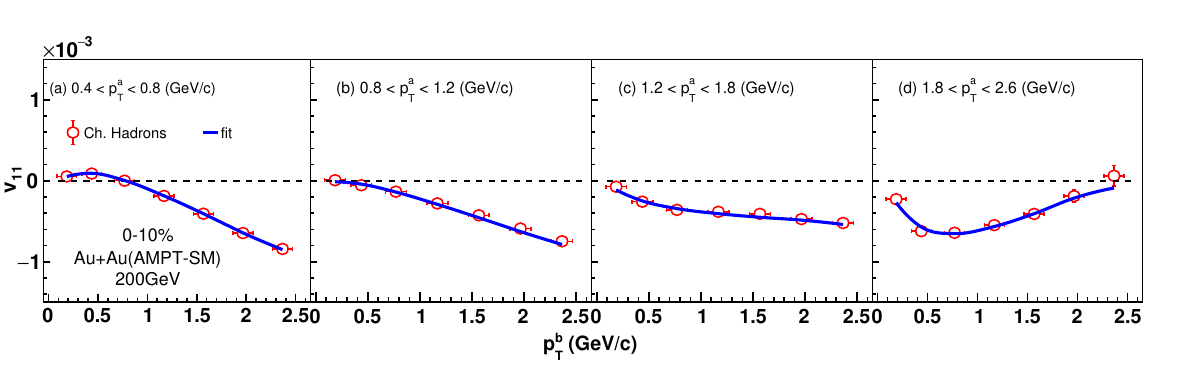}
\caption{ $v_{11}$ vs. $p_{T}^{\text{b}}$ for several selections of $p_{T}^{\text{a}}$  for 
0-10\% central Au+Au collisions at $\mathrm{\sqrt{s_{_{NN}}}} = 200$~GeV. The curve shows the 
result of the simultaneous fit with Eq.~\ref{eq_fit}.
}
\label{fig_method}
\end{figure*}

\section{The HIJING and AMPT Model}
The HIJING is a model used to simulate heavy-ion collisions. It is based on perturbative QCD and produces many small jet-like partons (minijets). These partons later form strings, which then break into hadrons using the Lund jet fragmentation model.
The model also includes the effect of parton shadowing by using a parameterized parton distribution function inside nuclei. However, HIJING does not include any mechanism to generate flow. Still, it is interesting to study how much the correlations between hadrons from minijets can contribute to the observed flow.\\

The AMPT model starts with the same initial conditions as HIJING. However, in the default mode of AMPT, the minijet partons are allowed to scatter (interact) with each other before they turn into hadrons through the Lund string fragmentation mechanism..
In the string-melting (SM) version of AMPT (called AMPT-SM),  strings are “melted” into partons. For example, mesons are converted into a quark–antiquark pair, and baryons into three quarks.
These quarks then interact through a parton cascade~\cite{zpc}. After the interactions stop, the partons combine to form hadrons through a process called coalescence. Because of these interactions among partons in AMPT-SM, the models can generate a significant amount of flow.  The subsequent hadronic matter interaction is described by a hadronic cascade, which is based on A Relativistic Transport (ART) model~\cite{art}.

In the AMPT model, the production probability of a quark--antiquark pair is proportional to $\exp(-\pi m_{T}^{2}/k)$, where $m_{T}$ is the transverse mass of the quark and $k$ is the string tension, which is approximately given by $k \propto 1/[b(2+a)]$. The string fragmentation parameters are chosen as $a = 0.55$ and $b = 0.15$, following Ref.~\cite{lund_para1}. The parton--parton scattering cross section is set to 3 mb, corresponding to a strong coupling constant of $\alpha_{S}=0.33$. Unless stated otherwise, the termination time of the hadronic cascade is fixed at 30~fm/$c$. To investigate the influence of hadronic rescattering on $v_1^{\mathrm{even}}$, the hadronic cascade termination time is varied in selected calculations.

\section{Analysis Method}
\label{sec:2pmethod}
The $v_{1}$ can be studied using the correlation function between two particles in relative azimuthal angle $\Delta\phi=\phi_{\mathrm{a}}-\phi_{\mathrm{b}}$~\cite{v1_even_4,v1_even_method_0,v1_even_method_1}.
The distribution of pairs in $\Delta\phi$ can be expanded into a Fourier series:
\begin{eqnarray}
\label{eq:2a}
 \frac{dN_{\mathrm{pairs}}}{d\Delta\phi} \propto 1+2\sum_{n,m =1}^{\infty}v_{n,m}(p_{T}^{\mathrm{a}},p_{T}^{\mathrm b}) \cos (n\phi_{\mathrm{a}}-m\phi_{\mathrm{b}})\;,
\end{eqnarray} 
where the coefficients $v_{n,m}$ are symmetric functions with respect to $p_{T}^{\mathrm{a}}$ and $p_{T}^{\mathrm b}$. 

The two-particle Fourier coefficients $v_{nn}$ are obtained from the correlation function as:
\begin{eqnarray}\label{vn}
 v_{nn} &=& \frac{\sum_{\Delta\phi} C_r(\Delta\phi)\cos(n \Delta\phi)}{\sum_{\Delta\phi}~C_r(\Delta\phi)},
\end{eqnarray}
where the correlation function $C_r(\Delta\phi)$ is calculated using
\begin{eqnarray}\label{corr_func}
 C_r(\Delta\phi) = \frac{(dN/d\Delta\phi)_{\text{same}}}{(dN/d\Delta\phi)_{\text{mixed}}}.
\end{eqnarray} 
  The term $(dN/d\Delta\phi)_{\text{same}}$ represents the azimuthal distribution of particle pairs from the same event, while $(dN/d\Delta\phi)_{\text{mixed}}$ corresponds to the azimuthal distribution of particle pairs formed by combining particles from different events with similar centrality. A  cut of $\Delta\eta = \eta_{a} - \eta_{b}$ is applied to the track pairs to reduce non-flow effects arising from short-range non-flow correlations.

The $v_{nn}$ values were then used to extract $v_{1}$ via a simultaneous fit of $v_{11}$ as 
a function of $p_{T}^{\text {b}}$ for several selections of  $p_{T}^{\text{a}}$ with Eq.~(\ref{eq_fit}),
\begin{equation}
\label{eq_fit}
v_{11}(p_{T}^{a},p_{T}^{b})  = v_{1}(p_{T}^{a})v_{1}(p_{T}^{b}) - K p_{T}^{a}p_{T}^{b}.
\end{equation}  

Here, $\mathrm{K \propto 1/(\langle N_{ch} \rangle \langle p_{T}^{2}\rangle)}$ takes into account  
the non-flow correlations induced by global momentum conservation,
$\mathrm{\langle N_{ch} \rangle}$ is the mean multiplicity and $\mathrm{\langle p_{T}^{2}\rangle}$ 
is  proportional to the variance of the transverse momentum over the full phase space. Figure~\ref{fig_method} shows $v_{11}$ vs. $p_{T}^{\text{b}}$ for several selections of $p_{T}^{\text{a}}$  for 0-10\% central Au+Au collisions at $\mathrm{\sqrt{s_{_{NN}}}} = 200$~GeV. The curve shows the result of the simultaneous fit with Eq.~\ref{eq_fit}.\\
In the absence of event-by-event fluctuations, the directed flow coefficient, $v_{1}$, is correlated with the impact-parameter direction and exhibits an odd dependence on pseudorapidity. However, fluctuations in the initial geometry can generate an additional rapidity-even component, $v_{1}^{\mathrm{even}}$.
The odd component of directed flow is antisymmetric in rapidity, satisfying $v_{1}^{\mathrm{odd}}(-y) = -v_{1}^{\mathrm{odd}}(y)$, and therefore averages to zero when integrated over a rapidity region symmetric about midrapidity. Consequently, the $v_{1}$ extracted using Eq.~(\ref{eq_fit}) within a rapidity-symmetric acceptance (e.g., $|y| < 1.0$) effectively corresponds to the rapidity-even component, $v_{1}^{\mathrm{even}}$.

\section{Results}
\begin{figure*}[ht]
\includegraphics[scale=0.4]{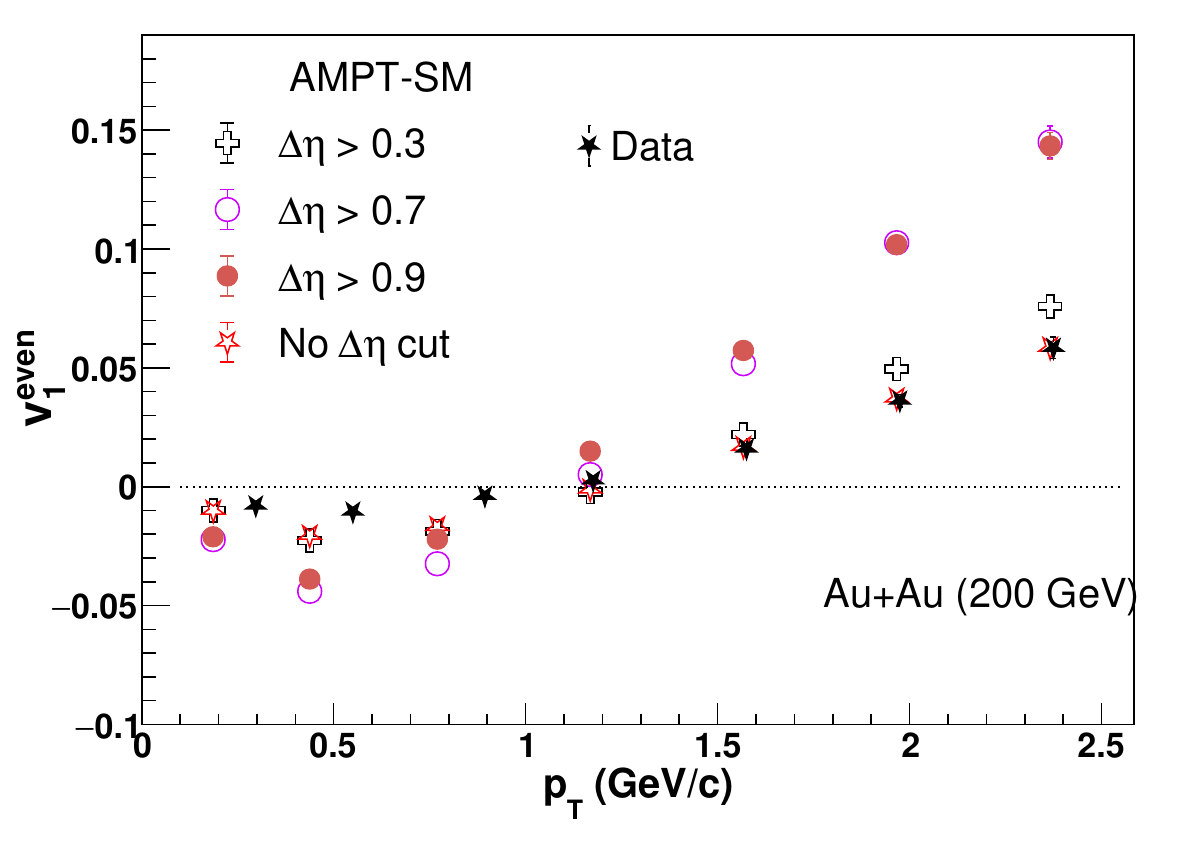}
\includegraphics[scale=0.38]{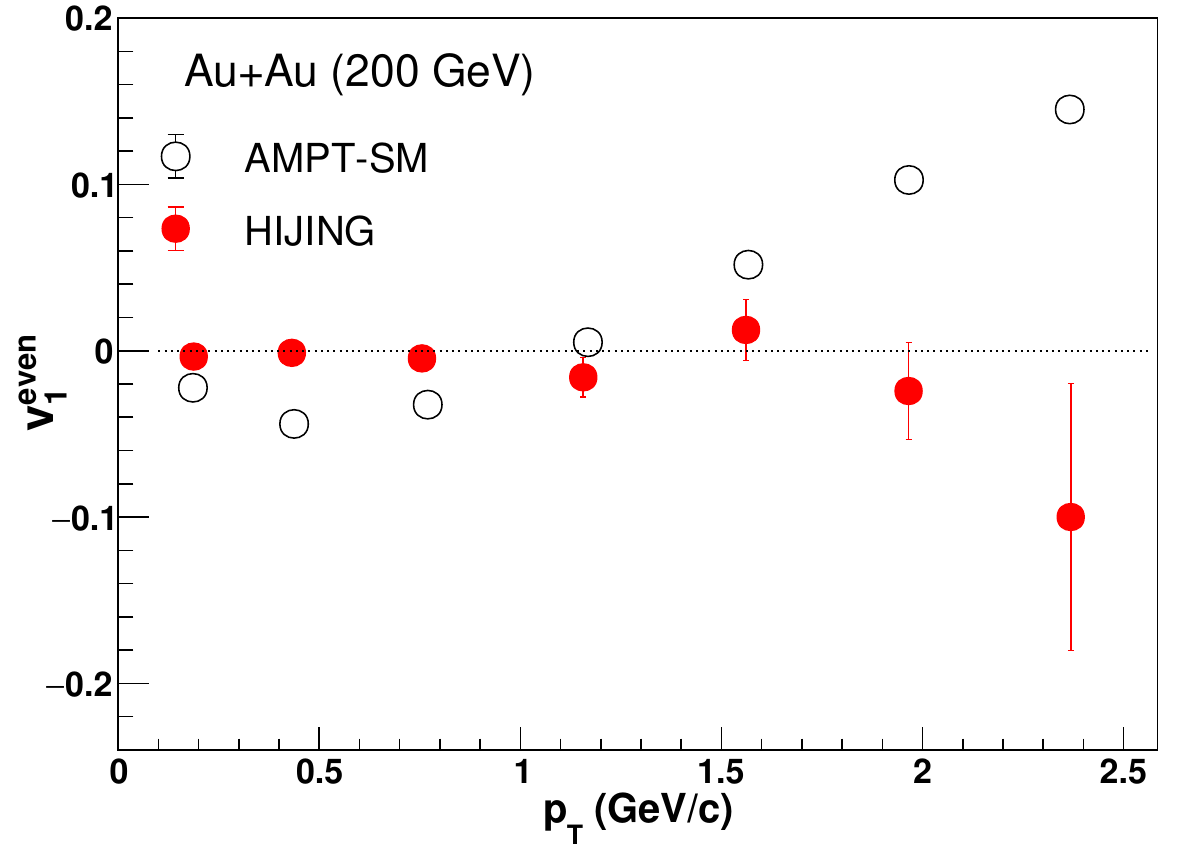}
\caption{Left Panel: The charged-hadron $v_1^{even}$ calculated using the AMPT model for 0-10\% Au+Au collisions at $\sqrt{s_{NN}} = 200$~GeV using  different pseudorapidity gaps, $|\Delta\eta| = 0, 0.7,$ and $0.9$ and within $|\eta|$ $<$ 1.0. Experimental measurement of  $v_1^{even}$  from Au+Au collisions at $\sqrt{s_{NN}} = 200$~GeV are also shown~\cite{v1_even_3}.
Right Panel: The charged-hadron $v_1^{even}$ is also calculated ($|\Delta\eta|$ $\geq$ 0.7) from HIJING model and combined with the calculation of the AMPT-SM model.
}
\label{fig_chg}
\end{figure*}

\begin{figure}[ht]
\includegraphics[scale=0.4]{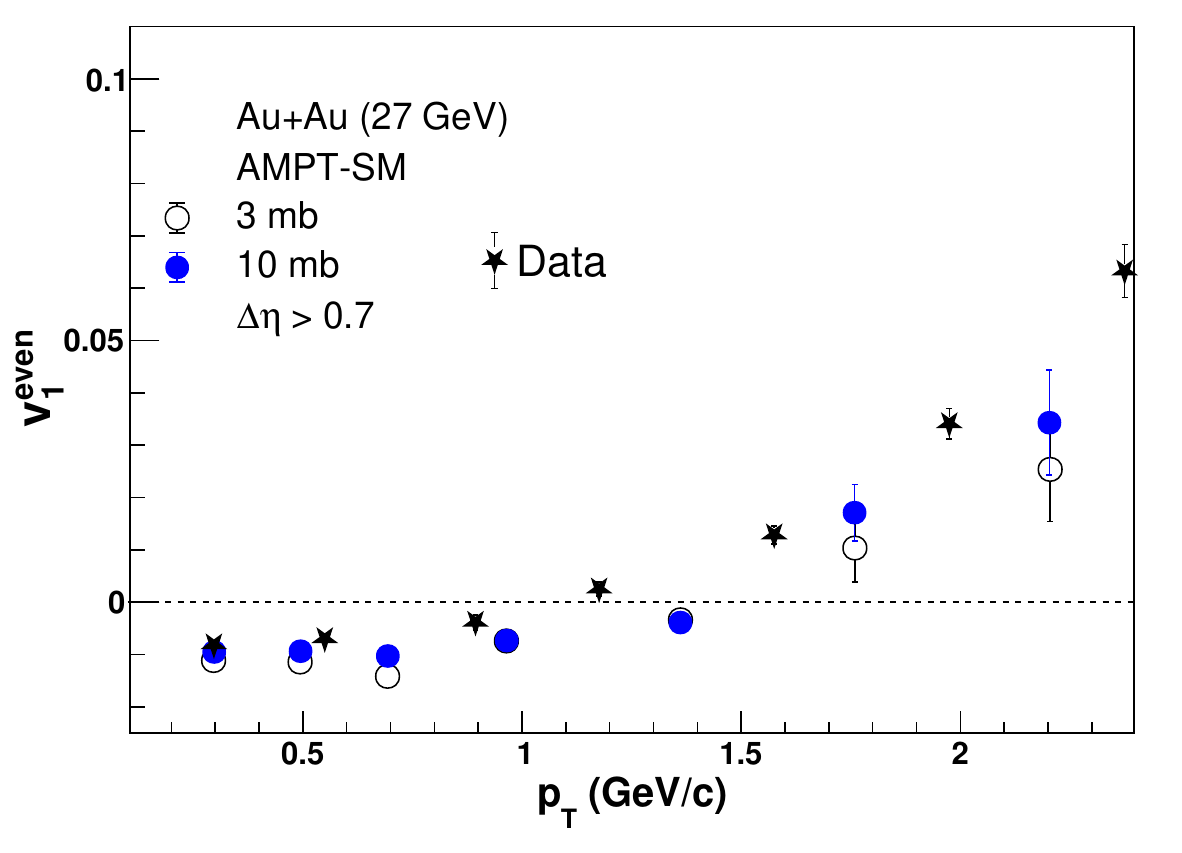}
\caption{The charged-hadron $v_1^{even}$ calculated using the AMPT model for 0-10\% Au+Au collisions at $\sqrt{s_{NN}} = 27$~GeV by varying parton-parton interaction cross-section. Experimental measurement of  $v_1^{even}$  from Au+Au collisions at $\sqrt{s_{NN}} = 27$~GeV are also shown~\cite{v1_even_3}.
}
\label{fig_chg_27}
\end{figure}

Fig.~\ref{fig_chg} (Left panel) shows the even component of charged-hadron $v_1$ calculated using the AMPT-SM model for Au+Au collisions at $\sqrt{s_{NN}} = 200$~GeV. The calculations are presented for different pseudorapidity gaps, $|\Delta\eta|$ $\geq$ 0, 0.7, and 0.9 and within $|\eta|$ $<$ 1.0. As observed, $v_{1}^{\mathrm{even}}$ varies with $|\Delta\eta|$, reflecting the reduction of non-flow effects such as resonance decays, Bose--Einstein correlations, and jets with increasing gap. In this paper, all results are presented for $|\Delta\eta| = 0.7$, since the difference between $|\Delta\eta| = 0.7$ and $0.9$ is not significant, while using $|\Delta\eta| = 0.7$ helps reduce statistical uncertainty.  Experimental measurements of $v_1^{\mathrm{even}}$ from Au+Au collisions at $\sqrt{s_{NN}} = 200$~GeV are also shown in Fig.~\ref{fig_chg} (left panel)~\cite{v1_even_3}. It can be seen that the AMPT model qualitatively reproduces the overall $p_T$ dependence of $v_1^{\mathrm{even}}$; however, it fails to provide a quantitative description of the data. The charged-hadron $v_1^{even}$ is also calculated from HIJING model and compared with the calculation of the AMPT model in Fig.~~\ref{fig_chg} (right panel).\\
 The HIJING model calculation shows that $v_{1}^{\mathrm{even}}$ is consistent with zero within statistical uncertainties, as expected, since HIJING does not generate collective flow. Any residual non-zero $v_{1}^{\mathrm{even}}$ in HIJING can be attributed to non-flow effects and global momentum conservation, which have been minimized using the method described in Section~\ref{sec:2pmethod}.\\
We have also checked the sensitivity of $v_1^{\mathrm{even}}$  on the parton-parton interaction cross-section and it is found  that $v_1^{\mathrm{even}}$  weakly depends on the parton--parton interaction cross section as shown in Fig.~\ref{fig_chg_27} for Au+Au collisions at $\sqrt{s_{NN}}$ = 27 GeV.  The consistency between data and AMPT model calculation is better at 27 GeV compared to that in 200 GeV.

\begin{figure*}[ht]
\includegraphics[scale=0.7]{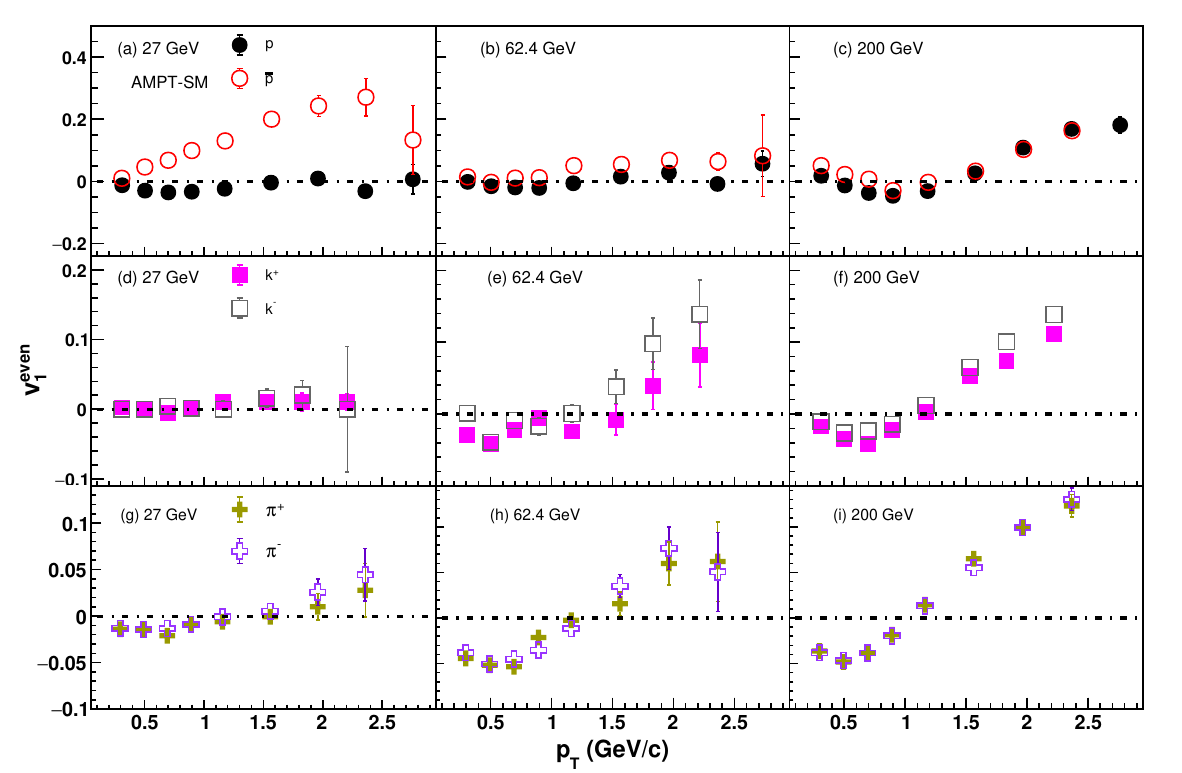}
\caption{The  $v_1^{even}$ of identified hadrons as a function of $p_{T}$ calculated using the AMPT-SM model for 0-10\% Au+Au collisions at $\sqrt{s_{NN}}$ = 27, 62.4 and  200 GeV. }
\label{fig_pid_pt}
\end{figure*}

Fig.~\ref{fig_pid_pt} presents the measurement of $v_{1}^{\mathrm{even}}$ for identified hadrons $\pi^{\pm}$, $K^{\pm}$, and $p(\bar{p})$ at mid-rapidity in Au+Au collisions at $\sqrt{s_{NN}} = 27$, $62.4$, and $200$~GeV using AMPT-SM. The results for particles and anti-particles are shown separately.
At the top RHIC energy, no significant difference is observed in $v_{1}^{\mathrm{even}}$ between particles and anti-particles. However, with decreasing center-of-mass energy, a pronounced difference between proton and anti-proton $v_{1}^{\mathrm{even}}$ is observed. In contrast, for mesons such as $\pi^{\pm}$ and $K^{\pm}$, no significant difference is observed between particles and their anti-particles.
The observed difference between proton and anti-proton is therefore likely related to baryon stopping effects at mid-rapidity in lower-energy collisions~\cite{trans_1,trans_2}.
This interpretation is consistent with previous observations of differences in $v_n$  between particles and anti-particles in the RHIC Beam Energy Scan program. The present measurement of $v_{1}^{\mathrm{even}}$, supported by AMPT model calculations, provides additional constraints on the physics of baryon transport and deposition at mid-rapidity.
\begin{figure*}[ht]
\includegraphics[scale=0.7]{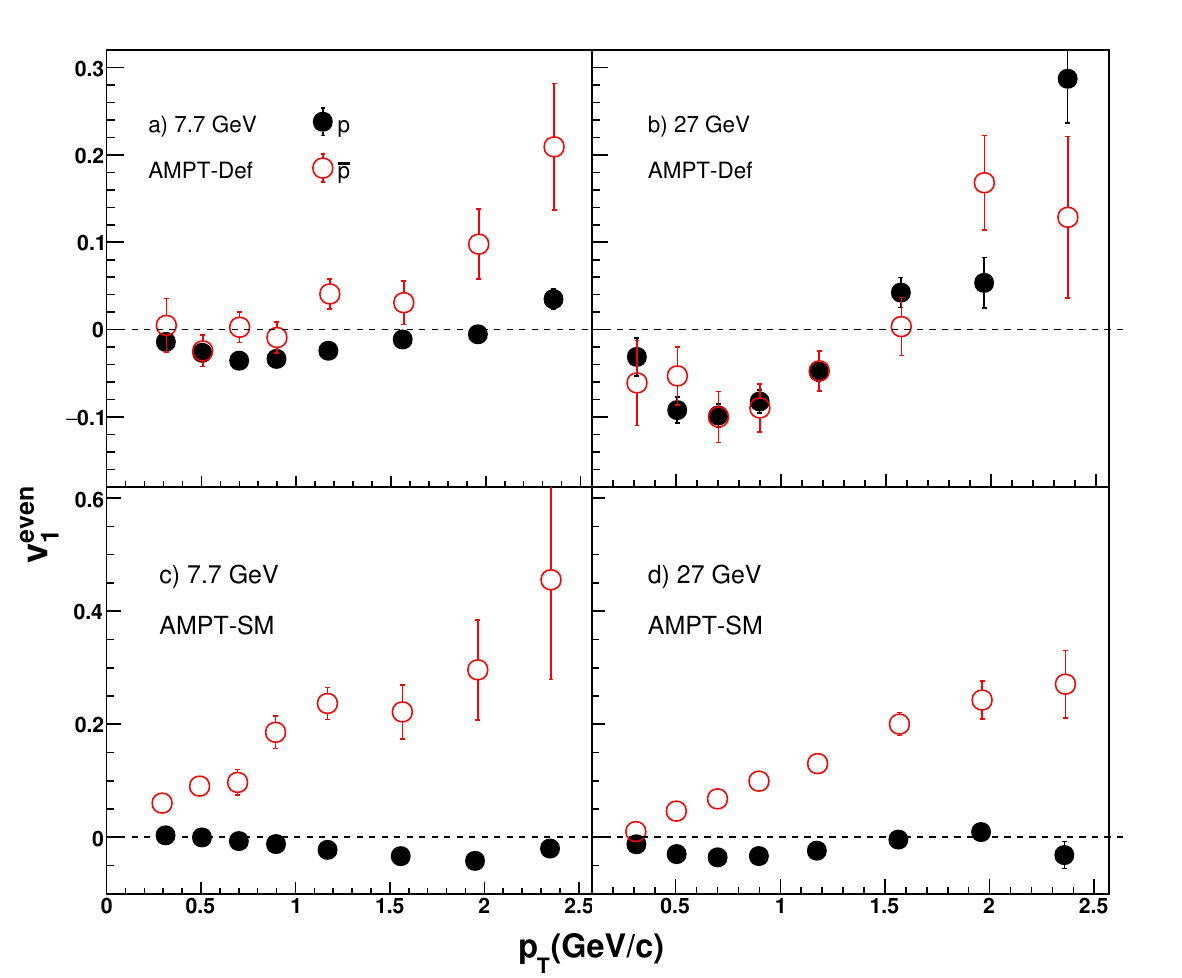}
\caption{ The $v_{1}^{even}$ of proton and anti-proton as a function $p_{T}$  for  0-10\% Au+Au collision at $\sqrt{s_{NN}} $= 27 and 7.7 GeV from AMPT-SM and AMPT-Def Model .}
\label{fig_v1_p_pbar_27gev}
\end{figure*}

\begin{figure*}[ht]
\includegraphics[scale=0.7]{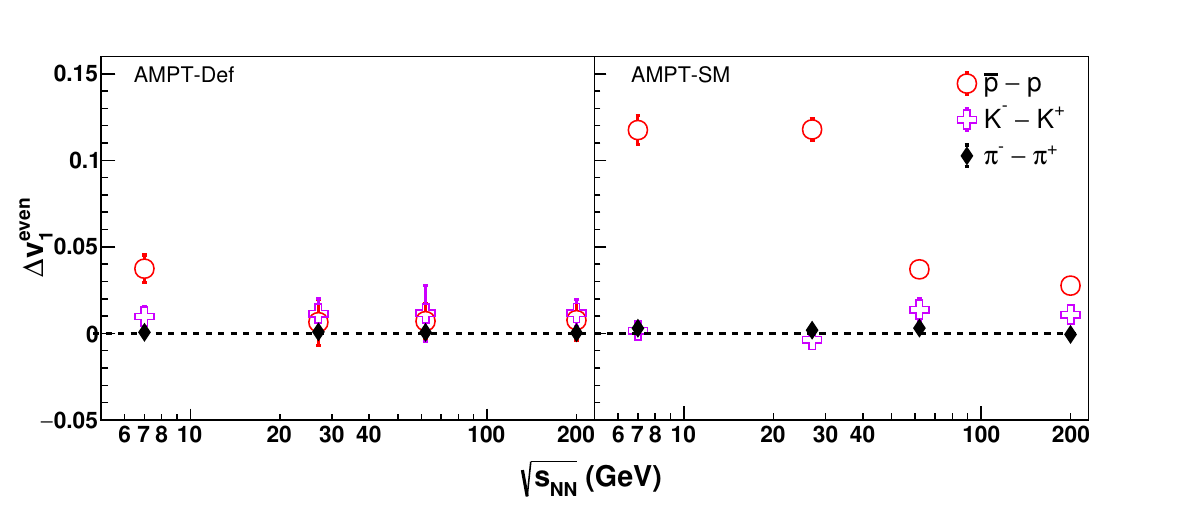}
\caption{ The difference in $v_{1}^{even}$ between particle and anti-particle integrated over $p_{T}$ are shown as a function of center-of-mass energy in 0-10\% collision centrality.}
\label{fig_delta_v1}
\end{figure*}

\begin{figure*}[ht]
\includegraphics[scale=0.7]{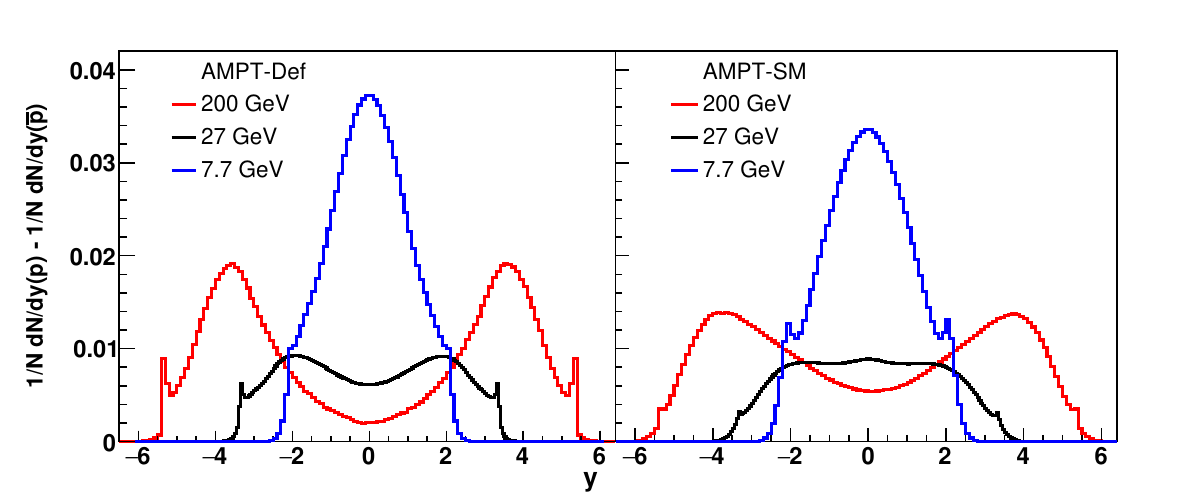}
\caption{Rapidity distribution of net-proton from AMPT-SM and AMPT-Default model is shown at 7.7,  27 and 200 GeV.}
\label{fig_net_p}
\end{figure*}

\begin{figure}[ht]
\includegraphics[scale=0.4]{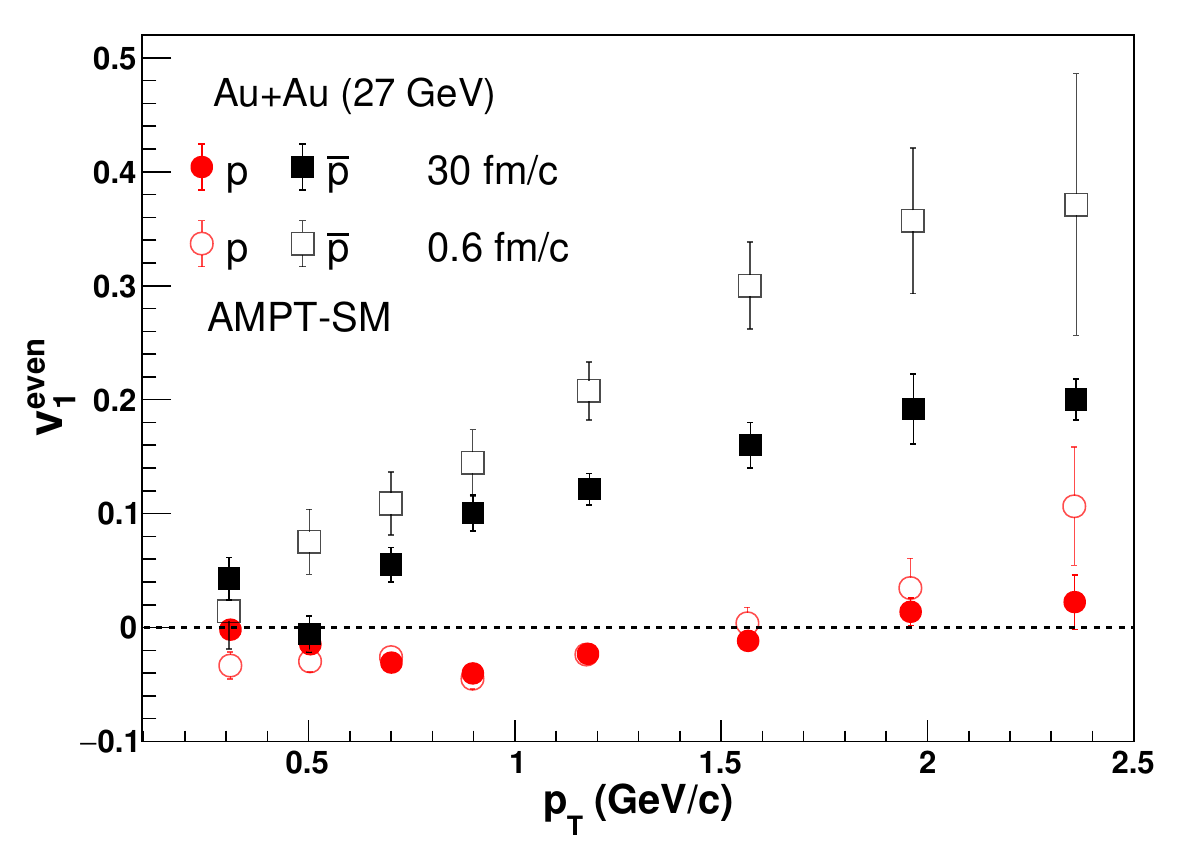}
\caption{ The $v_{1}^{even}$  of proton and anti-proton are shown with and without hadronic -rescattering using AMPT-SM model at 27 GeV for 0-10\% collision centrality.}
\label{fig_nt_v1}
\end{figure}

To further understand the mechanism responsible for the observed proton–antiproton $v_1^{even}$ difference, we compare results obtained from the AMPT-Default and AMPT-String Melting modes in Au+Au collisions at $\sqrt{s_{NN}} $= 27 and 7.7 GeV as shown in Fig.~\ref{fig_v1_p_pbar_27gev}. Interestingly, while a large proton–antiproton $v_1^{even}$ splitting is observed in the AMPT-SM calculations such a splitting  is not strong  in the AMPT-Default results. This contrast suggests that the mechanism responsible for the proton--antiproton flow splitting is determined not only by the amount of transported (stopped) quarks at midrapidity but also by how these transported quarks are distributed in rapidity. Therefore, the proton--antiproton $v_{1}$ splitting appears to be sensitive to both the magnitude of baryon stopping and the rapidity distribution of the stopped baryons. Furthermore, for antiprotons, strong annihilation during the hadronic evolution may further modify their directed flow~\cite{anhilation_1}.
Finally, The difference in $v_{1}^{\mathrm{even}}$ between particles and antiparticles, integrated over $p_{T}$ for 0--10\% central collisions, is shown in Fig.~\ref{fig_delta_v1}. The results from the AMPT string melting (AMPT-SM) model are presented in the right panel, while those from the AMPT-Default model are shown in the left panel. The two versions of AMPT differ mainly in their treatment of the partonic stage and hadronization. In the string melting mode, all excited strings are converted into constituent quarks and antiquarks, which undergo partonic scatterings and subsequently hadronize through quark coalescence. In contrast, the default mode retains the strings during the partonic evolution. Only minijet partons participate in the ZPC parton cascade, after which they recombine with their parent strings and hadronize through the Lund string fragmentation mechanism.\\
 In the AMPT-SM model, unlike pions and kaons, a clear separation between the $v_{1}^{\mathrm{even}}$ of protons and antiprotons is seen, and this separation becomes larger with decreasing beam energy. Such a difference is not large in the AMPT default model as compared to AMPT-SM. At lower energies, baryon stopping becomes increasingly important, leading to a larger fraction of transported quarks from the incoming nuclei reaching mid-rapidity.  As a result, protons containing transported quarks can develop a different flow pattern compared to antiprotons, which are mainly formed from produced quarks. This naturally leads to the observed proton--antiproton $v_{1}^{\mathrm{even}}$ splitting. In comparison to the AMPT-SM model, only a small difference between the proton and antiproton $v_{1}^{\mathrm{even}}$ is observed at $\sqrt{s_{NN}}=7.7$ GeV. To gain insight into this behavior, we compare the net-proton rapidity distributions obtained from the AMPT-SM and AMPT-Default models in Fig.~\ref{fig_net_p}. In the AMPT-Default model, the shape of the net-proton rapidity distribution at $\sqrt{s_{NN}}=27$ GeV remains similar to that at 200 GeV, whereas the AMPT-SM model exhibits a markedly different distribution at 27 GeV compared with 200 GeV. For 7.7 GeV, both AMPT-Def and AMPT-SM shows similar rapidity distribution of net-proton. These observations suggest that the proton--antiproton $v_{1}^{\mathrm{even}}$ splitting is sensitive not only to the degree of baryon stopping but also to the diffusion of stopped baryons in rapidity.
Finally, we investigated the effect of hadronic rescattering on the directed flow of protons and antiprotons by  varying the termination time of the hadronic cascade  t = 0.6 to 30 fm/$c$ in the AMPT-SM model. As shown in Fig.~\ref{fig_nt_v1}, the proton $v_{1}^{\mathrm{even}}$ remains nearly unchanged with and without the hadronic cascade. In contrast, the antiproton $v_{1}^{\mathrm{even}}$ is significantly reduced in the presence of the hadronic phase. This reduction is likely caused by antiproton annihilation during hadronic rescattering in the dense hadronic medium.\\
Besides the transported-quark mechanism and annihilation in hadronic medium, hadronic and partonic mean-field potentials have also been proposed as possible sources of the different collective dynamics of baryons and antibaryons in baryon-rich matter~\cite{mean_field_1,mean_field_2,mean_field_3,mean_field_4}. However, these effects are beyond the scope of the present study and have not been investigated in this work.

\section{Summary}
We study the rapidity-even directed flow, $v_{1}^{\mathrm{even}}$, of identified hadrons in Au+Au collisions at $\sqrt{s_{NN}}=27$, 62.4, and 200 GeV using the AMPT model. A significant proton–antiproton splitting in $v_{1}^{\mathrm{even}}$ develops toward lower beam energies, whereas pions and kaons exhibit no appreciable particle–antiparticle difference. The observed  proton–antiproton splitting   could be correlated with enhanced baryon stopping and transport to mid-rapidity at lower energies. Our results suggest that the proton–antiproton difference in $v_{1}^{\mathrm{even}}$ provides a sensitive probe of baryon stopping, baryon diffusion  and interaction in the hadronic phase. In general, this study highlights the importance of $v_{1}^{\mathrm{even}}$ measurements for identified hadrons at RHIC Beam Energy Scan energies.

\section{Appendix}
\begin{figure}[ht]
\includegraphics[scale=0.4]{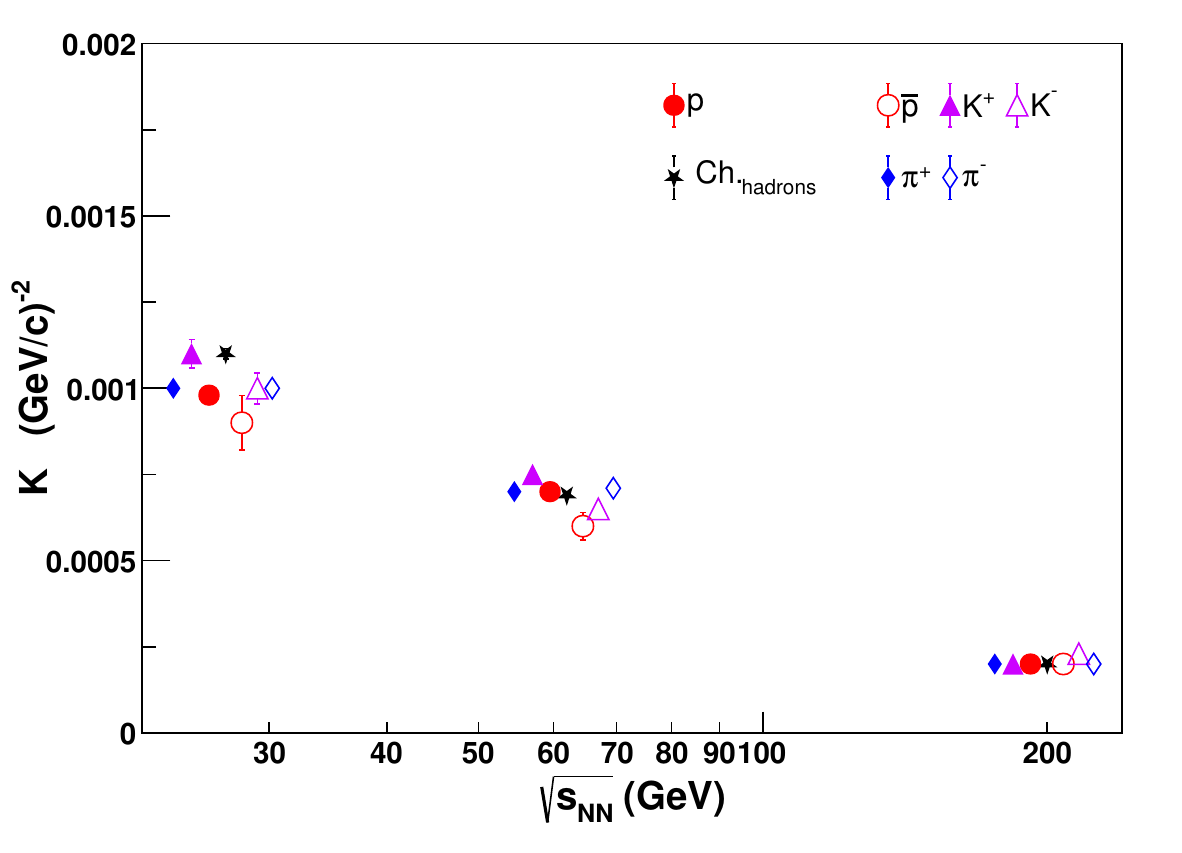}
\caption{ The values of $\mathrm{K \propto 1/(\langle N_{ch} \rangle \langle p_{T}^{2}\rangle)}$ obtained from the extraction of $v_{1}^{\mathrm{even}}$ using Eq.~\ref{eq_fit} for identified hadrons at $\sqrt{s_{NN}} = 27$, $62.4$, and $200$~GeV using AMPT-SM model. Data points are shifted horizontally for better visibility.}
\label{fig_k}
\end{figure}

In this section we have shown comparison of $\mathrm{K \propto 1/(\langle N_{ch} \rangle \langle p_{T}^{2}\rangle)}$ values obtained from simultaneous fit of $v_{11}$ as 
a function of $p_{T}^{\text {b}}$ for several selections of  $p_{T}^{\text{a}}$ with Eq.~(\ref{eq_fit}).
Fig.~\ref{fig_k} shows the values of $\mathrm{K \propto 1/(\langle N_{ch} \rangle \langle p_{T}^{2}\rangle)}$ obtained from the extraction of $v_{1}^{\mathrm{even}}$ using Eq.~\ref{eq_fit} for identified hadrons at different center-of-mass energies. For a given energy, the extracted $\mathrm{K}$ values do not vary significantly across different particle species.
This observation is consistent with the expectation that $\mathrm{K}$ depends only on the total event multiplicity $\langle N_{ch} \rangle$ and $\langle p_{T}^{2}\rangle$, where the latter is proportional to the variance of the transverse momentum over the full phase space. However, the magnitude of $\mathrm{K}$ is observed to increase with decreasing center-of-mass energy, reflecting the reduction in both multiplicity and $\langle p_{T}^{2}\rangle$ at lower energies.

\normalsize
\begin{thebibliography}{99}

\bibitem{flow1} J. Y. Ollitrault,  Phys. Rev. D 46, 229 (1992).
\bibitem{flow2} P. Huovinen, P. F. Kolb, U. Heinz, P. V. Ruuskanen, and S.A. Voloshin {\it et al.}, Phys. Lett. B 503, 58 (2001).
\bibitem{flow3} C. Shen and U. Heinz, Phys. Rev. C 85, 054902 (2012).
\bibitem{flow4} R. Snellings, New J. Phys. 13, 055008 (2011).

\bibitem{flow_method} A. M. Poskanzer, {\it et al.}, Phys. Rev. C 58, 1671 (1998).



\bibitem{v1_odd_0} H. Stoecker, Nucl. Phys. A, 750, 121 (2005).

\bibitem{v1_odd_1} P. Bozek and I. Wyskiel, Phys. Rev. C 81, 054902 (2010).

\bibitem{v1_odd_2} C. Alt, {\it et al.} (NA49 Collaboration), Phys. Rev. C 68, 034903 (2003).

\bibitem{v1_odd_3} J. Adams, {\it et al.} (STAR Collaboration), Phys. Rev. Lett. 92, 062301 (2004).

\bibitem{v1_odd_4} B. B. Back, {\it et al.} (PHOBOS Collaboration), Phys. Rev. Lett. 97, 012301 (2006).

\bibitem{v1_odd_5} Y. Nara, {\it et al.},  Phys. Rev. C, 94, 034906 (2016).


\bibitem{v3_paper} B. Alver and G. Roland, Phys. Rev. C 81 (2010) 054905 (2010).


\bibitem{v1_even_0} D. Teaney, Derek and L. Yan, Phys. Rev. C 83, 064904 (2011).
\bibitem{v1_even_1} M. Luzum, and J.-Y. Ollitrault, Phys. Rev. Lett. 106, 102301 (2011).
\bibitem{v1_even_2} B. Abelev, {\it et al.} (ALICE Collaboration), Phys. Rev. Lett. 111, 232302 (2013).
\bibitem{v1_even_3} J. Adams, {\it et al.} (STAR Collaboration) Phys. Lett. B 784, 26 (2018).
\bibitem{v1_even_4}  G. Aad,  {\it et al.} (ATLAS Collaboration), Phys. Rev. C 86, 014907 (2012).

\bibitem{v1e_p_ap_hyd} T. Parida, S. Chatterjee, arXiv: 2606.27156 (nucl-th)


\bibitem{trans_1} Y. Guo, F. Liu, and A. Tang, Phys. Rev. C 86, 044901 (2012).
\bibitem{trans_2} L. Adamczyk {\it et al.}, (STAR Collaboration)  Phys. Rev. Lett. 112, 162301 (2014)

\bibitem{anhilation_1} V. P. Konchakovski {\it et al.} ,  Phys. Rev. C 90, 014903 (2014).


\bibitem{mean_field_1}  C. M. Ko {\it et al.} , Nucl. Phys. A 928, 234 (2014).
\bibitem{mean_field_2}  P. Li {\it et al.},  Modern Physics Letters A 5, 2050289 (2020).
\bibitem{mean_field_3}   J. Xu  {\it et al.},  Phys. Rev. C 85, 041901(R) (2012).
\bibitem{mean_field_4} Y. Nara and A. Ohnishi, Phys. Rev. C 105, 014911 (2022). 

\bibitem{hijing} X. N. Wang and M. Gyulassy, Phys. Rev. D 44, 3501 (1991).

\bibitem{ampt} Zi-Wei Lin {\it et al.}, Phys. Rev. C 72, 064901 (2005).


\bibitem{zpc} B. Zhang, Comput. Phys. Commun. 109, 193 (1998).

\bibitem{art} B. A. Li and C. M. Ko, Phys. Rev. C 52, 2037 (1995).


\bibitem{lund_para1}   Z. W Lin , Phys. Rev. C 90, 014904 (2014).


\bibitem{v1_even_method_0} J. Jia, {\it et al.}, J. Phys. G 40, 105108 (2013).
\bibitem{v1_even_method_1} N. Borghini, {\it et al.}, Phys. Rev. C 62, 034902 (2000).

\end{thebibliography}
%

\end{document}